\documentclass[entropy,article,submit,oneauthor,pdftex,12pt,a4paper]{mdpi-nologo}
\lastpage{x}
\doinum{10.3390/------}
\pubvolume{xx}
\pubyear{2014}
\history{Received: xx / Accepted: xx / Published: xx}
\pdfoutput=1

\Title{Exact Probability Distribution versus Entropy}

\Author{Kerstin Andersson}

\address[1]{%
Department of Mathematics and Computer Science, Karlstad University, SE-651 88 Karlstad, Sweden\\
e-mail: Kerstin.Andersson@kau.se, telephone: +46-54-7001873}

\corres{e-mail: Kerstin.Andersson@kau.se, telephone: +46-54-7001873}

\abstract{
The problem addressed concerns the determination of the average
number of successive attempts of guessing a word of a certain length consisting
of letters with given probabilities of occurrence. Both first- and second-order
approximations to a natural language are considered. The guessing strategy used
is guessing words in decreasing order of probability. When word and alphabet
sizes are large, approximations are necessary in order to estimate the number of
guesses. Several kinds of approximations are discussed demonstrating moderate
requirements concerning both memory and CPU time. When considering realistic sizes
of alphabets and words (100) the number of guesses can be estimated within minutes
with reasonable accuracy (a few percent). For many probability distributions the
density of the logarithm of probability products is close to a normal distribution.
For those cases it is possible to derive an analytical expression for the average
number of guesses. The proportion of guesses needed on average compared to the
total number decreases almost exponentially with the word length. The leading term
in an asymptotic expansion can be used to estimate the number of guesses for large
word lengths. Comparisons with analytical lower bounds and entropy expressions are
also provided.
}


\begin{document}


\section{Introduction}

This work has been inspired by problems addressed in the field of
computer security, where the attacking of, e.g., password systems
is an important issue. In a brute-force attack the password,
for instance, can be broken in a worst-case time proportional to
the size of the search space and on average a time half of that.
However, if it is assumed that some words are more probable than
others, the words can be ordered in the search space in decreasing order
of probability. The number of guesses can then be drastically reduced.
Properties of the average number of successive guesses have been discussed
in detail by Pliam \cite{Pliam98}, who introduces the word guesswork to
denote this quantity. Further, Lundin {\it et al.} \cite{Lundin06} discuss
confidentiality measures related to guesswork.

As will be demonstrated below the calculation of the guesswork may require
a substantial amount of computational efforts, especially if the search
space is large. Therefore lower bounds, which are easy to calculate, have
been provided by several authors, e.g., Arikan \cite{Arikan96} and
Massey \cite{Massey94}.
Lower and upper bounds are provided by Pliam \cite{Pliam98}, but they involve
similar calculations as those needed for the guesswork itself and may
therefore be of less practical use.

In this paper numerical approaches are suggested for evaluating the average
number of successive guesses (guesswork) required for correctly guessing a
word from a given language. The guessing strategy used is guessing words
in decreasing order of probability. This is a continuation of investigations
presented elsewhere \cite{Andersson12}. In Section~\ref{lang} the languages used in this paper
are presented together with the corresponding expressions for the guesswork
and entropy. The reason for considering entropy here depends on the prevalent
use of entropy instead of guesswork in applications due to its simpler
determination. In Section~\ref{numEval} approximate numerical estimations
of guesswork are discussed and in Section~\ref{res} the results for some
probability distributions are given. Finally, in Section~\ref{conclusion}
the conclusions of the investigations presented in the paper are summarized.

\section{Languages}
\label{lang}

A language is a set of strings and a string is a finite sequence of
symbols from a given alphabet. Consider a stochastic variable $X$ belonging
to a state space $\mathcal{X} = \{x_1, x_2, \ldots, x_n\}$, where the probability
distribution is given by $p_X(x) = Pr(X=x)$. Introduce the short-hand notation
$p_i = p_X(x_i)$, where $\sum_{i=1}^n p_i = 1$.
In the following the state space $\mathcal{X}$ and its size $n$ are
considered as an alphabet with a certain number of symbols. Words are formed
by combining symbols to strings. From $n$ symbols it is possible to form $n^m$
different words of length $m$. Shannon introduced various orders of approximations to a
natural language, where the zero-order approximation is obtained by choosing
all letters independently and with the same probability. In the first-order
approximation the complexity is increased by choosing the letters according to their
probability of occurrence in the natural language. In zero- and first-order
approximation the strings thus consist of independent and identically-distributed
(iid) random variables. For higher-order approximations the variables are no longer
independent \cite{Shannon48}.

\subsection{Zero-order Approximation}
\label{zero}

In a zero-order approximation all symbols in the alphabet (of size $n$) have the
same probability of occurrence ($p_i = 1 / n, \forall i = 1,\ldots, n$).
The average number of guesses $G_0$ required to correctly guess a word of length $m$
is given by
\begin{eqnarray}
  G_0(X_1,\ldots,X_m) = \sum_{i=1}^{n^m} \left (\frac{1}{n}\right )^m i 
       = (n^m+1)/2,
  \label{gw0}
\end{eqnarray}
where $X_1,\ldots,X_m\in\mathcal{X}$. The entropy $H_0$ of a word of length $m$
is given by \cite{Shannon48}
\begin{eqnarray}
  H_0(X_1,\ldots,X_m) = \sum_{i=1}^{n^m} \left (\frac{1}{n}\right )^m \log_b n^m
      = m H_0(X_1),
  \label{e0} 
\end{eqnarray}
where $H_0(X_1)=\log_b n$ and $b$ is the base of the logarithm used.
The average number of guesses grows exponentially with the size of the word, while
the entropy grows linearly with the size of the word. This is in accordance with
the definition of entropy, since it should be an extensive property growing
linearly with the size of the system.

In a zero-order approximation the relation between guesswork and entropy is
$G_0(X_1,\ldots,X_m) = (b^{H_0(X_1,\ldots,X_m)} + 1)/2$. This relationship between
guesswork and entropy is true
in zero-order approximation, but not necessarily so using higher-order approximations,
which has been demonstrated by several authors (see, e.g., \cite{Pliam98}
and \cite{Malone05}). These authors strongly argue against the use of entropy
in the estimation of the number of required guesses.

\subsection{First-order Approximation}
\label{first}

In a first-order approximation the symbols in the alphabet (of size $n$) do not
necessarily have
the same probability of occurrence. Assume the symbols are ordered in decreasing
order of probability ($p_1\ge p_2\ge \ldots \ge p_n$). In first order the symbols in
a word are considered as stochastically independent and then the most probable word
(of a given length) would consist of only $x_1$. The most improbable word,
on the other hand, would consist of only $x_n$. The average number of
guesses $G_1$ required for making the correct guess of a word of length $m$ is
given by the summation:

\begin{equation}
  G_1(X_1,\ldots,X_m) = \sum_{i_1=1,\ldots, i_m=1}^n 
  p_{i_1}\cdots p_{i_m} g(i_1,\ldots,i_m), 
  \label{gw1}
\end{equation}
where the function $g(i_1,\ldots,i_m)$ represents the number of
guesses, one guess for the most probable word, two guesses for the second
most probable word and $n^m$ guesses for the most improbable word, etc.
The entropy $H_1$ of a word of length $m$ is given by \cite{Shannon48}
\begin{eqnarray}
  H_1(X_1,\ldots,X_m)&=&\sum_{i_1=1,\ldots, i_m=1}^n p_{i_1}\cdots p_{i_m} 
  \log_b\left(\frac{1}{p_{i_1}\cdots p_{i_m}}\right)\nonumber\\
  &=& m\sum_{i=1}^n p_i \log_b\left(\frac{1}{p_i}\right) = mH_1(X_1),
  \label{e1}
\end{eqnarray}
where $b$ is the base of the logarithm used.
The calculation of Equation~\ref{gw1} is more complicated than Equation~\ref{e1}
since it requires that the products of probabilities ($p_{i_1}\cdots p_{i_m}$)
are sorted in decreasing order. Such a procedure can be realized only for
a moderate size of $n^m$ due to both storage and CPU time requirements. For larger
values of $n^m$ approximate methods have to be used in order to get an estimate of
the summation. Lower bounds of the guesswork, that are easy to calculate, have been
provided by Massey \cite{Massey94} and Arikan \cite{Arikan96}. Massey demonstrates that

\begin{equation}
  G_1(X_1,\ldots,X_m) \ge \frac{1}{4}b^{H_1(X_1,\ldots,X_m)} + 1
  \label{gwMas}
\end{equation}
and Arikan that

\begin{equation}
  G_1(X_1,\ldots,X_m) \ge \frac{1}{1+m\ln n}\left [ \sum_{i=1}^{n} \sqrt{p_i}\right]^{2m}.
  \label{gwAri}
\end{equation}

\begin{figure}[htb]
  \centering
  \begin{minipage}[c]{0.9\textwidth}
    \centering
    \includegraphics[trim = 5mm 15mm 30mm 185mm, clip = true, angle=0, width=1.0\textwidth]{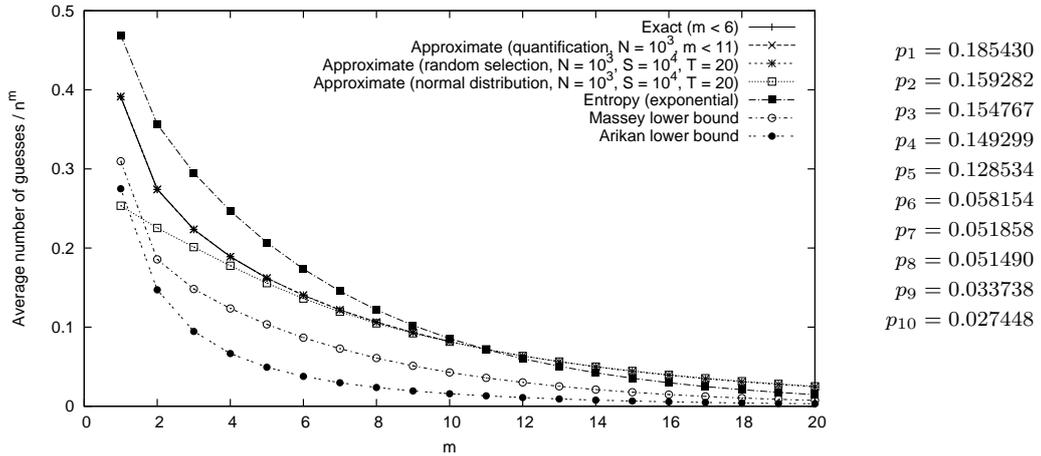}
  \end{minipage}
  \hspace{-4.0cm}
  \begin{minipage}[c]{0.2\textwidth}
    \centering
    {\scriptsize
    \begin{tabular}{r}
    $p_1 = 0.185430$\\
    $p_2 = 0.159282$\\
    $p_3 = 0.154767$\\
    $p_4 = 0.149299$\\
    $p_5 = 0.128534$\\
    $p_6 = 0.058154$\\
    $p_7 = 0.051858$\\
    $p_8 = 0.051490$\\
    $p_9 = 0.033738$\\
    $p_{10} = 0.027448$\\
    \\
    \\
    \end{tabular}
    }
  \end{minipage}
  \caption{The quotient of the average and the maximum number of
                  guesses of words of size $m$ for the randomly chosen
                  probability distribution given to the right ($n=10$).}
  \label{comp}
\end{figure}

In Figure~\ref{comp} the exact value of the guesswork for correctly guessing a
word of size $m < 6$ using an alphabet of size 10 (with the randomly chosen
probability distribution given in the figure) is displayed. The lower bounds
provided by Equations~\ref{gwMas} and~\ref{gwAri} are given for word sizes
$m \le 20$. For comparison, the exponential entropy expression

\begin{equation}
  \frac{1}{2}\left [b^{H_1(X_1,\ldots,X_m)} + 1\right ],
  \label{gwEnt}
\end{equation}
which is identical to guesswork in zero order, is given for word sizes $m \le 20$.
For word sizes $m < 6$ Equation~\ref{gwEnt} clearly overstimates the exact
value of the guesswork. In fact Pliam has shown that it is possible to construct
probability distributions that make guesswork differ arbitrariliy much from
Equation~\ref{gwEnt} \cite{Pliam98}. In Section~\ref{numEval} approximate numerical
evaluations of guesswork are discussed.

\subsection{Second-order Approximation}
\label{second}

In a second-order approximation the variables are no longer independent. Consider
two jointly distributed stochastic variables $X, Y \in \mathcal{X}$, then the
conditional probability distribution of $Y$ given $X$ is given by $p_Y(y|X=x) = 
Pr(Y=y|X=x)$. Introduce the short-hand notation $P_{ij} = p_Y(x_j|X=x_i)$, the
probability that symbol $x_j$ follows symbol $x_i$.
$P$ is an $n\times n$ matrix, where $n$ is the size of the alphabet and the
sum of the elements in each row is one. The probability of occurrence of each
symbol in the alphabet, $p_i$, can easily be obtained from matrix $P$ using
the two equations $(P^T - I)p = 0$ and $|p| = 1$, where $p$ is a vector of
length $n$ with elements $p_i$.

The guesswork $G_2$, i.e., the average number of guesses required for making the
correct guess of a word of length $m$  using an alphabet of size $n$ is
given by

\begin{equation}
  G_2(X_1,\ldots,X_m) = \sum_{i_1=1,\ldots, i_m=1}^n 
  p_{i_1}P_{i_1i_2}\cdots P_{i_{m-1}i_m} g(i_1,\ldots,i_m), 
  \label{gw2}
\end{equation}
where the function $g(i_1,\ldots,i_m)$ is the same as the one in Equation~\ref{gw1}.
The entropy $H_2$ of a word of length $m$ is given by \cite{Cover06}
\begin{eqnarray}
  \label{ent}
  H_2(X_1,\ldots,X_m)
  &=& -\sum_{i_1=1,\ldots, i_m=1}^n 
  p_{i_1}P_{i_1i_2}\cdots P_{i_{m-1}i_m}
  \log_b(p_{i_1}P_{i_1i_2}\cdots P_{i_{m-1}i_m}) \nonumber\\
  &=& -\sum_{i_1=1}^np_{i_1}\log_b(p_{i_1}) -\sum_{i_1=1}^np_{i_1}\sum_{i_2=1}^nP_{i_1i_2}\log_b(P_{i_1i_2})\nonumber\\
  &&- \sum_{i_1=1}^np_{i_1}\sum_{i_2=1}^nP_{i_1i_2}\sum_{i_3=1}^nP_{i_2i_3}\log_b(P_{i_2i_3}) - \ldots\nonumber \\
  &&- \sum_{i_1=1}^np_{i_1}\sum_{i_2=1}^nP_{i_1i_2}\cdots\sum_{i_m=1}^nP_{i_{m-1}i_m}\log_b(P_{i_{m-1}i_m})\nonumber\\
  &=& \sum_{i=1}^{m} H_2(X_i|X_{i-1}, \ldots , X_1),
  \label{e2}
\end{eqnarray}
where $b$ is the base of the logarithm used. In Section~\ref{res}
the value of Equation~\ref{gw2} will be compared to the value of Equation~\ref{gwEnt}
(with $H_1$ replaced by $H_2$) for a given probability distribution.

\section{Numerical Evaluation of Guesswork}
\label{numEval}

In this section a number of approaches will be given in order to evaluate the guesswork.

\subsection{Quantification}
\label{quant}

One simple procedure for numerically estimating Equation~\ref{gw1} and in addition
reducing the storage requirements is to split the range $\left [\log(\frac{1}{p_1}), 
\log(\frac{1}{p_n})\right ]$ into $N$ equal pieces of size $\Delta = 
\frac{1}{N}\log(\frac{p_1}{p_n})$, where a larger value of $N$ gives a better
estimate. The range $\left [\log(\frac{1}{p_1^m}), \log(\frac{1}{p_n^m})\right ]$ will
consequently be split into $m\cdot N$ equal pieces of size $\Delta$. Instead of
sorting the products $p_{i_1}p_{i_2}\cdots p_{i_m}$ they simply have to be evaluated
and brought into one of the $m\cdot N$ subranges. When the number of products in
each subrange has been determined an estimate of Equation~\ref{gw1} can be made,
giving
\begin{eqnarray}
  G_1(X_1,\ldots,X_m) &\approx& \sum_{j=1}^{mN} c_j\left [ C_j + \frac{1}{2}(c_j+1)\right ]P_j\nonumber\\
                      &=& G_1^Q(X_1,\ldots,X_m; N), 
  \label{gw1Approx}
\end{eqnarray}
where $c_j$ is the number of probability products in subrange $j$, $C_j = \sum_{k=1}^{j-1}c_k$
and $\log(P_j^{-1})$ is the middle value of subrange $j$. By instead using the boundary values
of the subranges, lower and upper bounds of the numerically estimated guesswork
can be given as

\begin{equation}
  \label{errQ1}
  G_1(X_1,\ldots,X_m) \in [Q_1^{-1}, Q_1]\cdot G_1^Q(X_1,\ldots,X_m; N),
\end{equation}
where $Q_1 = (p_1/p_n)^{1/2N}$. Here the short-hand notation $Q_1$ is used instead of
the more correct notation $Q_1(X_1,\ldots,X_m; N)$ in order to increase the transparency
of Equation~\ref{errQ1}.

By introducing the density of products $d_i = c_i/(\Delta n^m)$ the summations in
Equation~\ref{gw1Approx} can be replaced by integrals for large values of $N$, giving

\begin{equation}
  G_1(X_1,\ldots,X_m) \approx n^m (np_1)^m\int_{0}^{mN\Delta} dxb^{-x}d(x)\int_0^x dy d(y).
  \label{gw1Approx2}
\end{equation}
Equation~\ref{gw1Approx2} will be of importance in Section~\ref{norm}, where a
normal distribution approximation of the density of products is discussed.

The method of quantification can be used in both first- and second-order approximation.
However, since it is less obvious in second order which is the smallest and largest
value of the product of probabilities $p_{i_1}P_{i_1i_2}\cdots P_{i_{m-1}i_m}$
a lower bound of $\min(p_i)\cdot \min(P_{ij})^{m-1}$ and an upper bound of
$\max(p_i)\cdot \max(P_{ij})^{m-1}$ can be used instead to determine the range of possible
values. When determining $\min(P_{ij})$ only non-zero values are considered.
In second order a similar expression as the one in Equation~\ref{errQ1}
can be used for estimating the guesswork, namely

\begin{equation}
  \label{errQ2}
  G_2(X_1,\ldots,X_m) \in [Q_2^{-1}, Q_2]\cdot G_2^Q(X_1,\ldots,X_m; N),
\end{equation}
where $Q_2 = \left [\frac{\max(p_i)}{\min(p_i)}\right ]^{1/2mN}\cdot 
\left [\frac{\max(P_{ij})}{\min(P_{ij})}\right ]^{(m-1)/2mN}$
and $G_2^Q$ is given by Equation~\ref{gw1Approx} using the
values given above as interval limits for probability products.
Here the short-hand notation $Q_2$ is used instead of the more correct notation
$Q_2(X_1,\ldots,X_m; N)$ in order to increase the transparency of Equation~\ref{errQ2}.

\subsection{Random selection}
\label{rand}

The storage and CPU time requirements using the strategy in Section~\ref{quant} for
calculating the guesswork are of {\it O}($m\cdot N$) and {\it O}($m\cdot n^m$),
respectively. One simple modification for decreasing the time requirements
is to reduce the number of probability products formed. Instead of calculating
all $n^m$ different products a smaller number of randomly chosen probability products
is used and brought into the $m\cdot N$ subranges. The smaller number has
been determined to be proportional to $m$, i.e., equal to $m\cdot S$, where
$S$ is a parameter whose value has to be chosen. After normalization,
where the number of products in each subrange is multiplied by the factor
$n^m/(m\cdot S)$, the strategy is identical to the one in Section~\ref{quant}.

By not using all $n^m$ different probability products another error is introduced. This
error can be estimated by repeating the random selection calculations a number
of times (given by $T$). Through these calculations an average value ($G_i^R$)
and a standard deviation ($s_i^R$) can be estimated (where $i=1$ or 2). A 99~\%
confidence interval for $G_i^Q$ is then given as

\begin{equation}
   \label{errR}
   G_i^Q \in [1-R_i, 1+R_i]\cdot G_i^R,
\end{equation}
where $R_i = \lambda_{0.01/2}\cdot s_i^R/ (G_i^R\cdot \sqrt{T})$ and
$\lambda_{0.01/2}=2.58$ (the quantile function of the normal distribution) \cite{Box05}.
In Equation~\ref{errR} all parameters have been excluded in order to increase
the transparency of the equation.

\subsection{Normal distribution}
\label{norm}

\begin{figure}[h!t]
  \begin{center}
    \includegraphics[trim = 0mm 15mm 30mm 185mm, clip = true, angle=0, width=1.0\textwidth]{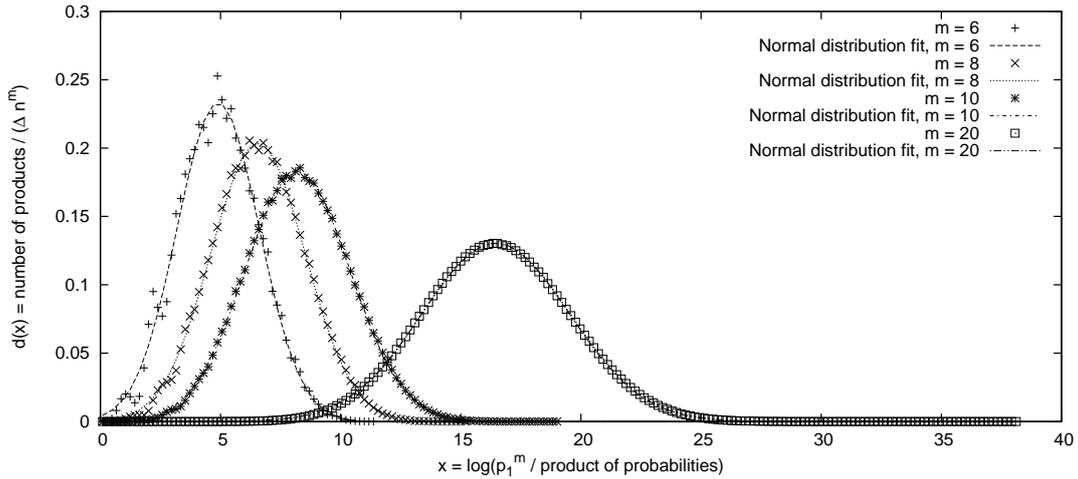}
    \caption{Density of logarithm of products of probabilities for a randomized
    probability distribution (given by Figure~\ref{comp}) for $n=10$ and $N=10$.
    Random selection with $S=10^8$ is used for $m=20$. Base $e$ is adopted. The
    average value is $\mu_1 m$ and the standard deviation $\sigma_1 \sqrt{m}$,
    where $\mu_1 = 0.824535$ and $\sigma_1 = 0.678331$.}
 \label{dens}
 \end{center}
\end{figure}

Another interesting approach is given by the central limit theorem in
probability theory, which roughly states that the mean of a large number
of independent stochastic variables is approximately normally distributed.
In Figure~\ref{dens} the density of the logarithm of products of probabilities
for the randomized probability distribution given in Figure~\ref{comp} is displayed.
The density fits nicely to a normal distribution, with a better fit for larger
values of $m$ (the number of independent stochastic variables). As expected the 
average value is proportional to $m$ and the standard deviation to $\sqrt{m}$. 
Denote the proportionality constants as $\mu_1$ and $\sigma_1$, respectively.

\begin{figure}[h!t]
  \begin{center}
    \includegraphics[trim = 15mm 15mm 45mm 190mm, clip = true, angle=0, width=0.9\textwidth]{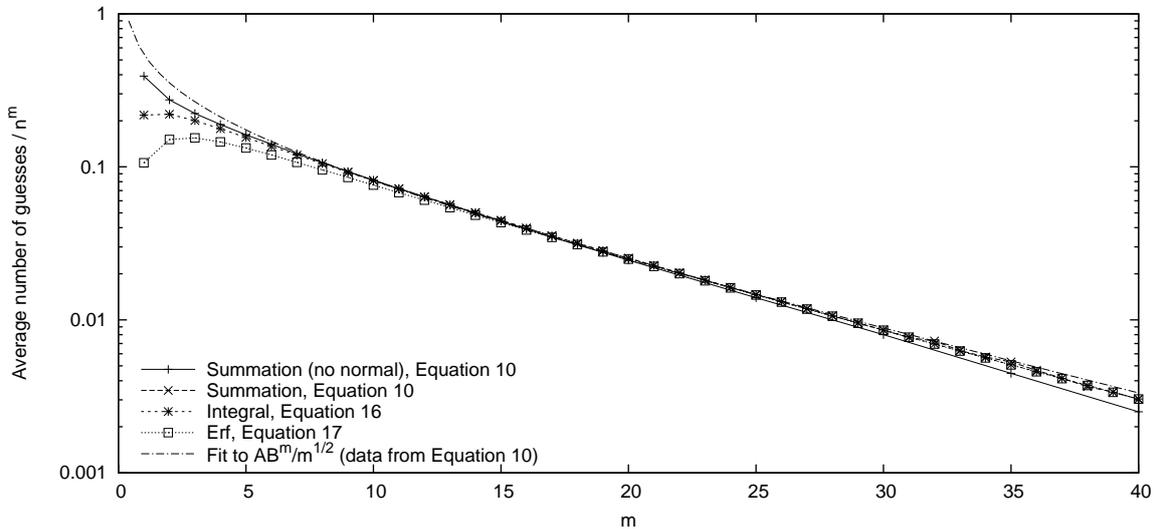}
    \caption{The guesswork for the randomized probability distribution given
    in Figure~\ref{comp} ($n=10$). For one summation the true density of logarithm 
    of probabilities is used ($N=10$, $S=10000$ and $T=20$) and otherwise a normal
    distribution is used (with the data given in Figure~\ref{dens} and $N=10$ for 
    Equation~\ref{gw1Approx}).}
 \label{gw1normal}
 \end{center}
\end{figure}

For large values of $n$ and $m$ and for most probability distributions
$\{p_1,\ldots,p_n\}$ it can be assumed (according to the central limit theorem)
that the logarithm of products of probabilities will be normally distributed.
The parameters of the normal distribution (the average value and the standard
deviation) can be estimated from a sample of a small number of random products
of probabilities (considerably smaller than required in the method described
in Section~\ref{rand}). The normal distribution is used to estimate the number of
probability products in each subrange, otherwise the strategy is identical to
the one in Section~\ref{quant}. When approximating the density of logarithms of
products by a normal distribution

\begin{equation}
  \label{normDist}
  N(x; \mu, \sigma^2) =\frac{1}{\sigma \sqrt{2\pi}}\cdot e^{-(x-\mu)^2/2\sigma^2},
\end{equation}
where $\mu$ is the average value and $\sigma$ the standard deviation,
Equation~\ref{gw1Approx2} can be expressed as
\begin{eqnarray}
  G_1(X_1,\ldots,X_m) \approx n^m \left (np_1 e^{-(2\mu_1 - \sigma_1^2)/2}\right )^m\nonumber\\
      \cdot\int_{0}^{\sqrt{m}N\Delta} dxN(x; (\mu_1-\sigma_1^2)\sqrt{m}, \sigma_1^2)
      \int_0^x dy N(y; \mu_1\sqrt{m},\sigma_1^2),
  \label{gw1Approx3}
\end{eqnarray}
where base $e$ has been adopted. The factor $e^{-x}$ (representing a product of pobabilities)
in Equation~\ref{gw1Approx2} cause a left shift of the normal distribution. This requires that also the tails
of the normal distribution are accurate in order for this approximation to be valid.
To make an error estimate for this kind of approximation is hard. However, if the
density of logarithm of probability products resembles a normal distribution also
at its tails then an error estimate similar to Equation~\ref{errR} can be made. Further,
the distance between the peaks of the two normal distributions in Equation~\ref{gw1Approx3}
is increasing for increasing values of $m$ resulting in decreasing values of the integral.
In fact it can be shown that Equation~\ref{gw1Approx3} can be further approximated as
\begin{eqnarray}
  G_1(X_1,\ldots,X_m) \approx n^m \left (np_1 e^{-(2\mu_1 - \sigma_1^2)/2}\right )^m
      \cdot\frac{1}{4}\left [ 2\textrm{erf}\left (\frac{\mu_1\sqrt{m}}{\sigma_1\sqrt{2}}\right ) - 1 
      - \textrm{erf}\left (\frac{\sigma_1\sqrt{m}}{2}\right )^2\right ]
  \label{gw1Approx4}
\end{eqnarray}
for large values of $m$\footnote{The integral $\int_0^pe^{-x^2}\textrm{erf}(p-x) = 
\frac{\sqrt{\pi}}{2}\left [ \textrm{erf}(\frac{p}{\sqrt{2}})\right ]^2$ from
Reference~\cite{Dorsogna} has been used.}. In Figure~\ref{gw1normal}
a comparison of Equations~\ref{gw1Approx} (with both the true and a normal distribution), \ref{gw1Approx3}
and~\ref{gw1Approx4} is given. The three expressions with a normal distribution give
the same value except for small values of $m$. The expression with the true distribution
resembles the others for moderate values of $m$. For $m>30$, however, they start to deviate
(see Section~\ref{res1}).

By using an asymptotic expansion of the error function\footnote{$\textrm{erf}(x) = 
1-\frac{e^{-x^2}}{x\sqrt{\pi}} \left [ 1 + \sum_{n=1}^\infty (-1)^n\frac{1\cdot 3\cdot 5\cdots 
(2n-1)}{(2x^2)^2}\right ]$ for large $x$.} it can be shown that the leading term
of Equation~\ref{gw1Approx4} is
\begin{eqnarray}
   n^m \cdot \frac{1}{\sigma_1\sqrt{\pi}}\cdot\left (np_1 e^{-(\mu_1 - \sigma_1^2/4)}\right )^m 
   \cdot \frac{1}{\sqrt{m}}
   \label{leading}
\end{eqnarray}
if $\mu_1/(\sigma_1\sqrt{2}) > \sigma_1/2$. Thus the leading term is of the form
$n^m\cdot A\cdot B^m\cdot m^{-1/2}$, where $A$ and $B$ are constants for the given
probability distribution. The result of fitting the data from Equation~\ref{gw1Approx}
(using a normal distribution) to such an expression is displayed in Figure~\ref{gw1normal}.
The results will be further discussed in the following section.

\section{Results}
\label{res}

In this section two probability distributions will be discussed. First, the distribution
given in Figure~\ref{comp} is investigated in more detail in Section~\ref{res1} and
second, the English language is considered in Section~\ref{res2}.

\subsection{Random probability distribution}
\label{res1}

In Figure~\ref{comp} the average number of guesses required for correctly guessing
a word of size $m$ using an alphabet of size 10 is given using various techniques.
First, the exact solution is given (for $m < 6$). Second, three approximate solutions
(as discussed in Section~\ref{numEval}) are given (quantification using all probability
products could be performed within reasonable time limits only for $m < 11$). Third,
an estimate based on entropy (Equation~\ref{gwEnt}) is provided. Fourth, lower bounds
derived by Massey (Equation~\ref{gwMas}) and Arikan (Equation~\ref{gwAri}) are included.

As is illustrated in Figure~\ref{comp} the approximate techniques of quantification and
random selection may provide accurate estimates of guesswork (with reasonable amount of
storage and CPU time). The third approximate technique (using a normal distribution) is
demonstrating accurate estimates for large values of $m$ ($>6$) in accordance with the
central limit theorem. By using a fitting procedure for values in the range $9 \le m \le 40$ an
approximate expression for guesswork is given by $G_1/n^m\approx 0.592\cdot 0.920^m\cdot m^{-1/2}$
(see Figure~\ref{gw1normal}). By evaluating the leading term of Equation~\ref{gw1Approx4}
(see Equation~\ref{leading}) the expression $0.832\cdot 0.912^m\cdot m^{-1/2}$ is obtained.

However, as is shown in Figure~\ref{gw1normal}, for large
values of $m$ the result given by a normal distribution deviates from that given by the
true distribution. When rescaling the $x$ axis in Figure~\ref{dens}, so that the range of
$x$ values is the same for all $m$ values the distribution becomes sharper for increasing
values of $m$. For sharper distributions the fitting to a normal distribution is more
sensitive. Small changes to the parameters $\mu$ and $\sigma$ may give large changes in
the value of guesswork.

The exponential entropy expression overestimates guesswork for small values of $m$ ($<11$)
and underestimates it for large values. The lower bound of Massey is closer to the exact
value than the lower bound of Arikan. However, both of the lower bounds underestimate
the number of guesses by an order of magnitude for $m=20$.

\subsubsection*{Error estimates}

Using the data in Figure~\ref{comp} and Equation~\ref{errQ1} the exact value can be determined
to be in the interval $[0.999, 1.001]\cdot G_1^Q$, i.e., the error using quantification is
about 0.1~\%. The additional error of using random selection (see Equation~\ref{errR}) is
determined to be between 0.26 and 0.56~\% (depending on the $m$ value) to a certainty of 99~\%.
The error due to random selection in the normal distribution appproximation is determined
to be between 0.30 and 0.63~\% (depending on the $m$ value) to a certainty of 99~\%. Observe
that this error does not include the fitness of a normal distribution to the density of the
logarithm of probability products.

\subsection{English language}
\label{res2}

\begin{figure}[htb]
  \centering
  \includegraphics[trim = 15mm 15mm 40mm 170mm, clip = true, angle=0, width=0.85\textwidth]{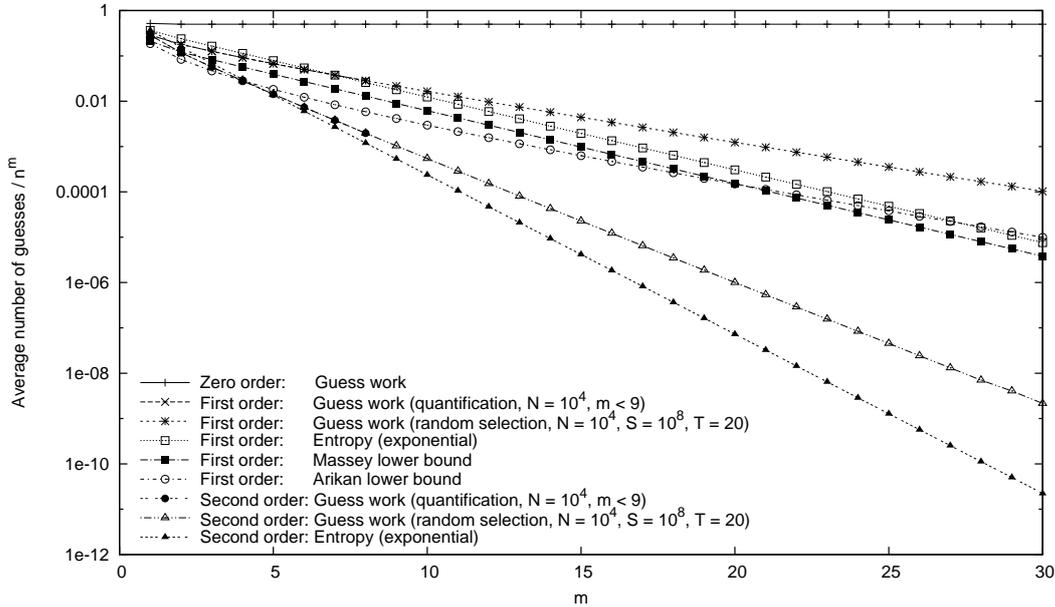}
  \caption{The quotient of the average and the maximum number of
                  guesses of words of size $m$ in the English language ($n=26$).}
  \label{order}
\end{figure}

While the probability distribution discussed in the previous section was randomly chosen,
the probability distribution considered here originates from the English language \cite{Digram}.
In Appendix~\ref{appendix} the English digram frequencies from Reference~\cite{Digram} are repeated.
In order to calculate the conditional probability distribution discussed in Section~\ref{second}
each row in the table in Appendix~\ref{appendix} has to be normalized. The probability distribution
for each letter in the English alphabet can be obtained by the procedures discussed
in Section~\ref{second}. In Figure~\ref{order} the average number of guesses required for
correctly guessing a word of size $m$ using the English alphabet of size 26 (with the
data given in Appendix~\ref{appendix}) is displayed. Guesswork has been numerically evaluated in zero,
first and second order. For comparison, estimates based on entropy (Equation~\ref{gwEnt})
are given in first and second order. In first order the lower bounds provided by Massey
(Equation~\ref{gwMas}) and Arikan (Equation~\ref{gwAri}) are included.

As is illustrated in Figure~\ref{order} all approaches display an exponential
behaviour (in first and second order) in accordance with Equations~\ref{gwMas},
\ref{gwAri}, \ref{gwEnt} and \ref{gw1Approx4}. A normal distribution was not
applied since it is not in agreement with the true distribution. It overestimates
guesswork by about an order of magnitude for $m=30$. However, it is possible to
make a fairly accurate fit of the guesswork data in first and second order to an
expression of the form
$n^m\cdot A\cdot B^m\cdot m^{-1/2}$ as was discussed in Section~\ref{norm}. By
using a fitting procedure for the guesswork graphs in Figure~\ref{order} for
$9\le m\le 30$ the average number of guesses of words in the English language
can be expressed according to the functions in Table~\ref{gwEst}. The deviation
between the true and estimated values (according to Table~\ref{gwEst}) is less
than 10~\% (except for the smallest $m$ values).

For both first and second order the
entropy ansatz underestimates the number of guesses required, for first order
by a factor of around 10 and for second order by a factor of around 100 for
word lengths of 30. Further, using the extra information provided by a
second-order approximation as compared to a first-order approximation reduces
the number of guesses by a factor of around $10^5$ for word lengths of 30.
The lower bounds of Massey (Equation~\ref{gwMas}) and Arikan (Equation~\ref{gwAri})
are underestimating the number of guesses by approximately the same amount as
the entropy expression for word lengths of 30.

\begin{table}
\caption{The average number of guesses of words of length $m$ in English
     divided by the maximum number.}
\label{gwEst}
\setlength{\tabcolsep}{3em}
\begin{center}
\begin{tabular}{|l|l|}
\hline
Order    & Expression \\
\hline
0        & $1/2$\\
1        & $0.481\cdot 0.801^m\cdot m^{-1/2}$\\
2        & $0.632\cdot 0.554^m\cdot m^{-1/2}$\\
\hline
\end{tabular}
\end{center}
\end{table}

\subsubsection*{Error estimates}
In first order the errors introduced by using quantification can be
calculated using Equation~\ref{errQ1}. Using the data in Figure~\ref{order}
the exact value can be determined to be in the interval
$[0.9998, 1.0002]\cdot G_1^Q$, where $G_1^Q$ is the approximate guess work
using quantification for first-order English. In second order Equation~\ref{errQ2}
and the data in Figure~\ref{order} make it possible to determine that the
exact value is in the interval $[0.9996, 1.0004]\cdot G_2^Q$.

Using the same procedure as in Section~\ref{res1} the error introduced
when randomly selecting probability products can be estimated. In first order
a 99~\% confidence interval for the guesswork is given by Equation~\ref{errR}
and by using the data in Figure~\ref{order} the error ($R_1$) is determined to be
between 0.006 and 0.05~\% (depending on the $m$ value). This should be added
to the error of 0.02~\% introduced by quantification. In second order exactly the
same procedure can be used and then the error is estimated to be in the range
0.006 and 6~\% (depending on the $m$ value). Again this error should be added
to the error of 0.04~\% introduced by quantification. The large error introduced
by random selection in second order for large $m$ values is due to the fact
that the fraction of probability products that are zero is larger for larger
$m$ values. By randomly selecting $S\cdot m$ probability products the number
of non-zero probability products is decreasing with an increasing value of $m$.
To increase the accuracy of the guesswork estimate in second order another
$m$ dependence of the number of selected probability products has to be chosen.

\section{Conclusion}
\label{conclusion}

In the paper it has been demonstrated that it is possible to estimate the average
number of guesses (guesswork) of a word with a given length numerically with
reasonable accuracy (of a couple of percent) for large alphabet sizes ($\approx 100$)
and word lengths ($\approx 100$) within minutes. Thus, a numerical estimate of
guesswork constitutes an alternative to, e.g., various entropy expressions.

For many probability distributions the density of the logarithm of probability products
is close to a normal distribution. For those cases it is possible to derive an analytical
expression for guesswork showing the functional dependence of the word length. The
proportion of guesses needed on average compared to the total number decreases almost
exponentially with the word length. The leading term in an asymptotic expansion of
guesswork has the form $n^m\cdot A\cdot B^m\cdot m^{-1/2}$, where $A$ and $B$ are constants
(however, different for different probability distributions), $n$ is the size of the
alphabet and $m$ is the word length. Such an expression can be determined for medium-sized
values of $m$, using some fitting procedure, and used with fairly good accuracy for large values of $m$.

In the paper the English language has been investigated. The average number of guesses
has been calculated numerically in both first and second order giving a reduction of the
number of guesses by a factor $10^5$ for word lengths of 30 when the extra information
provided by second order is included. A normal distribution of the logarithm of probability
products was not applied since it is not in agreement with the true distribution. It
overestimates guesswork by about an order of magnitude for word lengths of 30. Still
it is possible to find accurate expressions for guesswork ($0.481\cdot 0.801^m\cdot m^{-1/2}$
in first order and $0.632\cdot 0.554^m\cdot m^{-1/2}$ in second order) in agreement with
the true values (the deviation is less than 10~\% for word lengths of 30).

Comparison between guesswork and entropy expressions has been performed showing that
the entropy ansatz underestimates the number of guesses required, for first order
by a factor of around 10 and for second order by a factor of around 100 for English
words of length 30. Lower bounds of guesswork by Massey and Arikan have also been
investigated. They are underestimating the number of guesses by approximately the
same amount as the entropy expression for word lengths of 30.

\appendix
\section{English Digram Frequencies}
\label{appendix}

The information in the matrix below is used for creating matrix $P$ used in
Section~\ref{second} \cite{Digram}. After normalization of each row matrix $P$
is obtained.

\noindent
\begin{table}[hb]
\centering
\scalebox{0.52} {
\begin{tabular}{r|rrrrrrrrrrrrrrrrrrrrrrrrrr}
   &  A&  B&  C&  D&  E&  F&  G&  H&  I&  J&  K&  L&  M&  N&  O&  P&  Q&  R&  S&  T&  U&  V&  W&  X&  Y&  Z\\
\hline
A  &  1& 32& 39& 15&  0& 10& 18&  0& 16&  0& 10& 77& 18&172&  2& 31&  1&101& 67&124& 12& 24&  7&  0& 27&  1\\
B  &  8&  0&  0&  0& 58&  0&  0&  0&  6&  2&  0& 21&  1&  0& 11&  0&  0&  6&  5&  0& 25&  0&  0&  0& 19&  0\\
C  & 44&  0& 12&  0& 55&  1&  0& 46& 15&  0&  8& 16&  0&  0& 59&  1&  0&  7&  1& 38& 16&  0&  1&  0&  0&  0\\
D  & 45& 18&  4& 10& 39& 12&  2&  3& 57&  1&  0&  7&  9&  5& 37&  7&  1& 10& 32& 39&  8&  4&  9&  0&  6&  0\\
E  &131& 11& 64&107& 39& 23& 20& 15& 40&  1&  2& 46& 43&120& 46& 32& 14&154&145& 80&  7& 16& 41& 17& 17&  0\\
F  & 21&  2&  9&  1& 25& 14&  1&  6& 21&  1&  0& 10&  3&  2& 38&  3&  0&  4&  8& 42& 11&  1&  4&  0&  1&  0\\
G  & 11&  2&  1&  1& 32&  3&  1& 16& 10&  0&  0&  4&  1&  3& 23&  1&  0& 21&  7& 13&  8&  0&  2&  0&  1&  0\\
H  & 84&  1&  2&  1&251&  2&  0&  5& 72&  0&  0&  3&  1&  2& 46&  1&  0&  8&  3& 22&  2&  0&  7&  0&  1&  0\\
I  & 18&  7& 55& 16& 37& 27& 10&  0&  0&  0&  8& 39& 32&169& 63&  3&  0& 21&106& 88&  0& 14&  1&  1&  0&  4\\
J  &  0&  0&  0&  0&  2&  0&  0&  0&  0&  0&  0&  0&  0&  0&  4&  0&  0&  0&  0&  0&  4&  0&  0&  0&  0&  0\\
K  &  0&  0&  0&  0& 28&  0&  0&  0&  8&  0&  0&  0&  0&  3&  3&  0&  0&  0&  2&  1&  0&  0&  3&  0&  3&  0\\
L  & 34&  7&  8& 28& 72&  5&  1&  0& 57&  1&  3& 55&  4&  1& 28&  2&  2&  2& 12& 19&  8&  2&  5&  0& 47&  0\\
M  & 56&  9&  1&  2& 48&  0&  0&  1& 26&  0&  0&  0&  5&  3& 28& 16&  0&  0&  6&  6& 13&  0&  2&  0&  3&  0\\
N  & 54&  7& 31&118& 64&  8& 75&  9& 37&  3&  3& 10&  7&  9& 65&  7&  0&  5& 51&110& 12&  4& 15&  1& 14&  0\\
O  &  9& 18& 18& 16&  3& 94&  3&  3& 13&  0&  5& 17& 44&145& 23& 29&  0&113& 37& 53& 96& 13& 36&  0&  4&  2\\
P  & 21&  1&  0&  0& 40&  0&  0&  7&  8&  0&  0& 29&  0&  0& 28& 26&  0& 42&  3& 14&  7&  0&  1&  0&  2&  0\\
Q  &  0&  0&  0&  0&  0&  0&  0&  0&  0&  0&  0&  0&  0&  0&  0&  0&  0&  0&  0&  0& 20&  0&  0&  0&  0&  0\\
R  & 57&  4& 14& 16&148&  6&  6&  3& 77&  1& 11& 12& 15& 12& 54&  8&  0& 18& 39& 63&  6&  5& 10&  0& 17&  0\\
S  & 75& 13& 21&  6& 84& 13&  6& 30& 42&  0&  2&  6& 14& 19& 71& 24&  2&  6& 41&121& 30&  2& 27&  0&  4&  0\\
T  & 56& 14&  6&  9& 94&  5&  1&315&128&  0&  0& 12& 14&  8&111&  8&  0& 30& 32& 53& 22&  4& 16&  0& 21&  0\\
U  & 18&  5& 17& 11& 11&  1& 12&  2&  5&  0&  0& 28&  9& 33&  2& 17&  0& 49& 42& 45&  0&  0&  0&  1&  1&  1\\
V  & 15&  0&  0&  0& 53&  0&  0&  0& 19&  0&  0&  0&  0&  0&  6&  0&  0&  0&  0&  0&  0&  0&  0&  0&  0&  0\\
W  & 32&  0&  3&  4& 30&  1&  0& 48& 37&  0&  0&  4&  1& 10& 17&  2&  0&  1&  3&  6&  1&  1&  2&  0&  0&  0\\
X  &  3&  0&  5&  0&  1&  0&  0&  0&  4&  0&  0&  0&  0&  0&  1&  4&  0&  0&  0&  1&  1&  0&  0&  0&  0&  0\\
Y  & 11& 11& 10&  4& 12&  3&  5&  5& 18&  0&  0&  6&  4&  3& 28&  7&  0&  5& 17& 21&  1&  3& 14&  0&  0&  0\\
Z  &  0&  0&  0&  0&  5&  0&  0&  0&  2&  0&  0&  1&  0&  0&  0&  0&  0&  0&  0&  0&  0&  0&  0&  0&  0&  1\\
\end{tabular}
}
\end{table}



\acknowledgements{Acknowledgements}

The author would like to thank \AA ngpannef\"oreningens Forskningsstiftelse
for a travelling scholarship. The author would also like to thank for the
valuable remarks and comments received in connection with the conferences
CiE 2012 and UCNC 2012.












\bibliographystyle{mdpi}
\bibliography{bibfile}

\begin{thebibliography}{-------}
\providecommand{\natexlab}[1]{#1}

\bibitem[Pliam(1998)]{Pliam98}
Pliam, J.O.
\newblock The disparity between work and entropy in cryptology.
\newblock Theory of Cryptography Library, Record 98--24,  1998.

\bibitem[Lundin \em{et~al.}(2006)Lundin, Lindskog, Brunstr{\"o}m, and
  Fischer-H{\"u}bner]{Lundin06}
Lundin, R.; Lindskog, S.; Brunstr{\"o}m, A.; Fischer-H{\"u}bner, S.
\newblock Using guesswork as a measure for confidentiality of selectivly
  encrypted messages. In {\em Advances in Information Security}; Gollmann, D.;
  Massacci, F.; Yautsiukhin, A., Eds.; Springer: New York, NY,  2006; Vol.~23,
  pp. 173--184.

\bibitem[Arikan(1996)]{Arikan96}
Arikan, E.
\newblock An inequality on guessing and its application to sequential decoding.
\newblock {\em IEEE Transactions on Information Theory} {\bf 1996}, {\em
  42},~99--105.

\bibitem[Massey(1994)]{Massey94}
Massey, J.L.
\newblock Guessing and entropy.
\newblock {\em Proc. IEEE Int. Symp. Information Theory} {\bf 1994}, p. 204.

\bibitem[Andersson(2012)]{Andersson12}
Andersson, K.
\newblock Numerical evaluation of the average number of successive guesses.
\newblock  UCNC 2012; Durand-Lose, J.; Jonoska, N., Eds.; Springer: Heidelberg,
   2012; Vol. 7445, {\em LNCS}, p. 234.

\bibitem[Shannon(1948)]{Shannon48}
Shannon, C.E.
\newblock A mathmatical theory of communication.
\newblock {\em The Bell System Technical Journal} {\bf 1948}, {\em
  27},~379--423, 623--656.

\bibitem[Malone and Sullivan(2005)]{Malone05}
Malone, D.; Sullivan, W.
\newblock Guesswork is not a substitute for entropy.
\newblock  Proceedings of the Information Technology and Telecommunications
  Conference,  2005.

\bibitem[Cover and Thomas(2006)]{Cover06}
Cover, T.M.; Thomas, J.A.
\newblock {\em Elements of Information Theory}, 2nd. ed.; Wiley: Hoboken,
  2006.

\bibitem[Box \em{et~al.}(2005)Box, Hunter, and Hunter]{Box05}
Box, G.E.P.; Hunter, J.S.; Hunter, W.G.
\newblock {\em Statistics for experimenters: design, innovation and discovery},
  2nd. ed.; Wiley: Hoboken,  2005.

\bibitem[D'Orsogna(2013)]{Dorsogna}
D'Orsogna, M.R.
\newblock \url{http://mathworld.wolfram.com/Erf.html}.

\bibitem[Nicholl(2013)]{Digram}
Nicholl, J.
\newblock \url{http://jnicholl.org/Cryptanalysis/Data/DigramFrequencies.php}.

\end{thebibliography}

\end{document}